\documentclass{article}
\usepackage{amssymb}

\input{tcilatex}
\begin{document}

\author{Emilio Santos \\
Universidad de Cantabria. Santander. Spain. \and Email \TEXTsymbol{<}%
emilio.santos@unican.es\TEXTsymbol{>}.}
\title{Local model of entangled photon experiments compatible with quantum
predictions based on the reality of the vacuum fields. }
\date{September, 8, 2020 }
\maketitle

\begin{abstract}
Arguments are provided for the reality of the quantum vacuum fields. A
polarization correlation experiment with two maximally entangled photons
created by spontaneous parametric down-conversion is studied in the
Weyl-Wigner formalism, that reproduces the quantum predictions. An
interpretation is proposed in terms of stochastic processes assuming that
the quantum vacuum fields are real. This proves that local realism is
compatible with a violation of Bell inequalities, thus rebutting the claim
that it has been refuted by experiments. Entanglement appears as a
correlation between fluctuations of a signal field and vacuum fields.

Key words; local realism, Bell inequalities, entangled photons, Weyl-Wigner,
loopholes,vacuum fields
\end{abstract}

\section{ Introduction}

The purpose of this article is to rebut the standard wisdom that ``local
realism'' has been refuted by experiments\cite{Shalm},\cite{Giustina}
measuring polarization correlation of entangled photon pairs produced via
spontaneous parametric down-conversion in non-linear crystals. For the proof
I will present a local realistic model of the experiments resting on the
assumption that the quantum vacuum fields are real stochastic fields. In
section 2 I will provide arguments for the reality of the vacuum fields.
After that I will briefly recall the quantum interpretation of the
experiments, firstly using the common Hilbert-space formalism of quantum
theory applied to quantum optics in section 3. Then using the Weyl-Wigner
formalism, which is physically equivalent to the former, in section 4.
Finally in section 5 I will show that the latter formalism may be
interpreted in terms of random variables and stochastic processes thus
providing a local realistic model of the experiments. In this introductory
section I include a short discussion about the relevance of local realism
and the Bell inequalities.

\subsection{The conflict between quantum mechanics and local realism}

The interpretation of quantum mechanics has been the subject of continuous
debate from the very begining of the theory. Initially the question was
whether quantum mechanics is complete or not, but later the celebrated EPR
article\cite{EPR} showed an incompatibility between completeness and
locality (in the sense of relativistic causality). The authors supported
locality, something that Einstein strongly backed until his death\cite
{Einstein}. In 1964 Bell showed that locality may be empirically tested via
his celebrated inequalities\cite{Bell}, \cite{Brunner}. Indeed the standard
wisdom is that any violation of a Bell inequality refutes locality, meaning
that not all natural phenomena are compatible with relativistic causality.
Thus the results of the alleged loophole-free empirical tests\cite{Shalm},%
\cite{Giustina} has been interpreted as the ``death by experiment for local
realism'', this being the hypothesis that ``the world is made up of real
stuff, existing in space and changing only through local interactions ...
about the most intuitive scientific postulate imaginable''\cite{Wiseman}. In
the present article I prove that local realism has not been refuted if we
accept that the quantum vacuum fields are real.

\subsection{The Bell inequalities}

Bell defined as ``local hidden variables'' model, later named ``local
realistic'', any model of an experiment where the results of all correlation
measurements may be interpreted according to the formulas

\begin{eqnarray}
\left\langle A\right\rangle &=&\int \rho \left( \lambda \right) d\lambda
M_{A}\left( \lambda ,A\right) ,\left\langle B\right\rangle =\int \rho \left(
\lambda \right) d\lambda M_{B}\left( \lambda ,B\right) ,  \nonumber \\
\left\langle AB\right\rangle &=&\int \rho \left( \lambda \right) d\lambda
M_{A}\left( \lambda ,A\right) M_{B}\left( \lambda ,B\right) ,  \label{bell}
\end{eqnarray}
where $\lambda \in \Lambda $ is one or several random\ (``hidden'')
variables, $\left\langle A\right\rangle ,\left\langle B\right\rangle $ and $%
\left\langle AB\right\rangle $ being the expectation values of the results
of measuring the observables $A,B$ and their product $AB$, respectively.
Here we will consider that the observables correspond to detection, or not,
of some signals (e. g. photons) by two parties, say Alice and Bob, attaching
the values 1 or 0 to these two possibilities. In this case $\left\langle
A\right\rangle ,\left\langle B\right\rangle $ agree with the single and $%
\left\langle AB\right\rangle $ with the coincidence detection rates
respectively in typical experiments consisting in repeated trials. The
following mathematical conditions are assumed 
\begin{equation}
\rho \left( \lambda \right) \geq 0,\int \rho \left( \lambda \right) d\lambda
=1,M_{A}\left( \lambda ,A\right) \in \left\{ 0,1\right\} ,M_{B}\left(
\lambda ,B\right) \in \left\{ 0,1\right\} .  \label{bell1}
\end{equation}
A constraint of locality is included, namely $M_{A}\left( \lambda ,A\right) $
should be independent of the choice of the observable $B$, $M_{B}\left(
\lambda ,B\right) $ independent of $A,$ and $\rho \left( \lambda \right) $
independent of both $A$ and $B$. From these conditions it is possible to
derive empirically testable (Bell) inequalities\cite{CH}, \cite{Eberhard}.
The tests are most relevant if the measurements performed by Alice and Bob
are spacially separated in the sense of relativity theory.

For experiments measuring polarization correlation of photon pairs the
Clauser-Horne inequality\cite{CH} may be written

\begin{equation}
\left\langle \theta _{1}\right\rangle +\left\langle \phi _{1}\right\rangle
\geq \left\langle \theta _{1}\phi _{1}\right\rangle +\left\langle \theta
_{1}\phi _{2}\right\rangle +\left\langle \theta _{2}\phi _{1}\right\rangle
-\left\langle \theta _{2}\phi _{2}\right\rangle ,  \label{CH}
\end{equation}
where $\theta _{j}$ stands for the observable ``detection of a photon with
the Alice detector in front of a polarizer at angle $\theta _{j}$''.
Similarly $\phi _{k}$ for Bob detector. For simplicity in this paper I will
study the case of maximally entangled photons, although in the mentioned
experiments\cite{Shalm},\cite{Giustina} the photon pairs had partial
entanglement that made the experiments easier. I will present a local model
that predicts the following single and coincidence rates by Alice and Bob 
\begin{eqnarray}
\left\langle \theta _{j}\right\rangle &=&\left\langle \phi _{k}\right\rangle
=K,  \nonumber \\
\left\langle \theta _{j}\phi _{k}\right\rangle &=&K\cos ^{2}\left( \theta
_{j}-\phi _{k}\right) =\frac{1}{2}K\left[ 1+\cos \left( 2\theta _{j}-2\phi
_{k}\right) \right] ,  \label{pred}
\end{eqnarray}
where $K$ is a constant that depends on the particular experimental setup.
It is easy to check that the prediction eq.$\left( \ref{pred}\right) $
violates the inequality eq.$\left( \ref{CH}\right) $ for some choices of
angles. For instance the choice 
\[
\theta _{1}=\frac{\pi }{4},\phi _{1}=\frac{\pi }{8},\theta _{2}=0,\phi _{2}=%
\frac{3\pi }{8}, 
\]
violates the inequality eq.$\left( \ref{CH}\right) $ leading to 
\[
K\ngtr \frac{1}{2}\left( 1+\sqrt{2}\right) K\simeq 1.207K. 
\]
So far we have assumed ideal detectors, for real detectors the above
predicted single rates $\left\langle \theta _{j}\right\rangle $ and $%
\left\langle \phi _{k}\right\rangle $ should be multiplied times the
detection efficiencies $\eta _{A}$ and $\eta _{B},$ respectively, and the
coincidence rate $\left\langle \theta _{j}\phi _{k}\right\rangle $ times $%
\eta _{A}\eta _{B},$ whence the empirical violation of the inequality eq.$%
\left( \ref{CH}\right) $ would require high detection efficiencies, that is 
\[
\eta _{A}+\eta _{B}<(1+\sqrt{2})\eta _{A}\eta _{B}\Rightarrow \eta >0.828%
\text{ if }\eta _{A}=\eta _{B}=\eta . 
\]
Experiments with some non-maximal entanglement need only $\eta >2/3$ \cite
{Eberhard}, that was the reason for using such entanglement in the actual
experiments \cite{Shalm}, \cite{Giustina}.

\section{The assumption that the vacuum fields are real}

\subsection{The quantum vacuum in field theory}

The existence of some equilibrium radiation energy in space, even at zero
Kelvin, appears for the first time in Planck's second radiation theory of
1912. This zeropoint energy of the electromagnetic field (ZPF) was rejected
because it is divergent, although the consequences of its possible reality
were soon explored by several authors\cite{Milonnibook}. The ZPF reappeared
in 1927 when Dirac quantized the electromagnetic field via an expansion in
normal modes, that is plane waves in the case of free space. In fact the
Hamiltonian of the field may be written 
\begin{equation}
H=\frac{1}{2}
\rlap{\protect\rule[1.1ex]{.325em}{.1ex}}h%
\sum_{j}\omega _{j}\left( a_{j}+a_{j}^{\dagger }\right) ^{2}=\frac{1}{2} 
\rlap{\protect\rule[1.1ex]{.325em}{.1ex}}h%
\sum_{j}\omega _{j}(a_{j}^{2}+a_{j}^{\dagger 2}+2a_{j}^{\dagger }a_{j}+1),
\label{0}
\end{equation}
where $\omega _{j}$ is the frequency of a normal mode, $a_{j}$ and $%
a_{j}^{\dagger }$ being the annihilation and creation operators of photons
in that mode and we have taken the commutation rules into account in the
later equality. For every mode a single index $j$ is used that includes both
wavevector and polarization. The energy is given by the vacuum expectation
of the Hamiltonian, that is 
\begin{eqnarray}
\left\langle 0\left| H\right| 0\right\rangle &=&\frac{1}{2} 
\rlap{\protect\rule[1.1ex]{.325em}{.1ex}}h%
\sum_{j}\omega _{j}\left\langle 0\left| (a_{j}^{2}+a_{j}^{\dagger
2}+2a_{j}^{\dagger }a_{j}\right| 0\right\rangle +\frac{1}{2} 
\rlap{\protect\rule[1.1ex]{.325em}{.1ex}}h%
\sum_{j}\omega _{j}\left\langle 0\left| 1\right| 0\right\rangle  \nonumber \\
&=&\sum_{j}(\frac{1}{2}
\rlap{\protect\rule[1.1ex]{.325em}{.1ex}}h%
\omega _{j}),  \label{00}
\end{eqnarray}
the former expectation being nil. The result corresponds to a mean energy $%
\frac{1}{2}
\rlap{\protect\rule[1.1ex]{.325em}{.1ex}}h%
\omega _{j}$ per mode, whence the total energy density in space diverges
when we sum over all (infinitely many) normal modes.

The standard solution to the divergence problem is to remove the term that
contributes in eq.$\left( \ref{00}\right) $, a procedure which is known as
``normal ordering''. It consists of writing the annihilation operators to
the right, that is to assume that the correct Hamiltonian is not the former
expression in eq.$\left( \ref{0}\right) ,$ but the latter with unity
removed. It may be realized that the normal ordering is equivalent to
choosing the zero of energies at the level of the vacuum. It provides a
practical procedure useful in quantum-mechanical calculations, but it is not
a good solution in the opinion of many authors. They see it as an ``ad hoc''
assumption, just aimed at removing unpleasant divergences. For these authors
the ZPF is a logical consequence of quantization and the solution of the
divergence problem should come from a more natural mechanism. Furthermore it
has been shown that the assumption of reality of the ZPF combined with the
classical laws of electrodynamics allows explaining some phenomena usually
taken as purely quantal, an approach known as stochastic electrodynamics\cite
{dice}.

Around 1947 two new discoveries reinforced the hypothesis that the quantum
vacuum fields are real, namely the Lamb shift and the Casimir effect. Willis
Lamb observed an unexpected absorption of microwave radiation by atomic
hydrogen, that was soon explained in terms of the interaction of the atom
with the quantized electromagnetic field, that involves the vacuum radiation
(ZPF). Indeed Lamb claimed to be the discoverer of the ZPF by experiment.
Furthermore he wrote that ``photons are the quanta of the electromagnetic
field, but they are not particles''\cite{Lamb}. Lamb discovery led in a few
years to the development of quantum electrodynamics (QED), a theory that
allows predictions in spectacular agreement with experiments, and it was the
starting point for the whole theory of relativistic quantum fields. The
success of QED rests on renormalization techniques, namely assuming that
physical particles, e. g. electrons, are dressed with ``virtual fields''
making their physical mass and charge different from the bare quantities. In
my opinion these assumptions behind renormalization are actually a
reinforcement of the reality of the quantum vacuum fields, although people
avoid commitement with that conclusion using the word ``virtual'' as an
alternative to ``really existing''.

The Casimir effect\cite{Casimir} consists of the attraction between two
parallel perfectly conducting plates in vacuum. The force $F$ per unit area
depends on the distance $d$ between the plates, 
\begin{equation}
F=-\frac{\pi ^{2}
\rlap{\protect\rule[1.1ex]{.325em}{.1ex}}h%
c}{240d^{4}},  \label{Casimir}
\end{equation}
a force confirmed empirically\cite{Milonnibook}. The reason for the
attraction may be understood qualitatively as follows. In equilibrium the
electric field of the zeropoint radiation, ZPF, should be zero or normal to
any plate surface, otherwise a current would be produced. This fact
constrains the possible low frequency normal modes of the radiation, that is
those having wavelengths of the order the distance between plates or larger,
although the distribution of high frequency modes is barely modified by the
presence of the plates. Therefore attaching an energy $\frac{1}{2} 
\rlap{\protect\rule[1.1ex]{.325em}{.1ex}}h%
\omega _{j}$ to every mode, the total energy of the ZPF in space becomes a
function of the distance between plates and the derivative of that function
with respect to the distance leads to eq.$\left( \ref{Casimir}\right) .$
Actually the calculated energy of the ZPF diverges if we sum over all
radiation modes, including those with arbitrarily high frequency, but there
are regularization procedures that give the correct result\cite{Milonnibook}%
. They essentially subtract the field energy with the plates present minus
the energy with the plates removed. The physical picture of the phenomenon
is that the radiation pressures in both faces of each plate are different
and this is the reason for a net force on the plate. The Casimir effect is
currently considered the most strong argument for the reality of the quantum
vacuum fields. For us it is specially relevant because it provides an
example of the fact that the difference between the radiation arriving at
the two faces of a plate is what matters, rather than the total radiation
acting on the plate. We will make a similar assumption in the model of
detector to be proposed in section 5 below.

\subsection{The quantum vacuum in astrophysics and cosmology}

In laboratory physics, where gravity usually plays no role, the reality of
the quantum vacuum fields is not too relevant a question. In fact, their
possibly huge, or divergent, energy may be ignored choosing the zero of
energy at the level of the vacuum, that is using the normal ordering rule.
However this choice is no longer innocuous in the presence of gravity
because, according to relativity theory, energy gravitates whence a huge
energy would produce a huge gravitational field. Therefore the possible
existence of a vacuum energy is a relevant question in astrophysics and
cosmology. In the following my personal view is presented about the possible
rebuttal of objections to the reality of the vacuum fields.

From long ago the quantum vacuum has been related to the cosmological
constant, a term that Einstein introduced in general relativity in order to
make possible his 1917 model of universe. The reason for the relation is
that the vacuum, even if it possesses some energy density, $\rho $, should
be Lorentz invariant. Then it should exists also pressure, $P$, with the
equation of state $P=-\rho ,$ that is equivalent to a cosmological constant
term. Actually there was no empirical evidence for a cosmological constant
until 1999\cite{cosmolterm}, but even before that date many authors
speculated about the possibility that the quantum vacuum fields give rise to
a cosmological term. However there was a big problem, namely the vacuum
energy density appears to be infinite if no cut-off exists, and the only
natural cut-off seems to be at the Planck scale. This cut-off would give a
cosmological term about $10^{124}$ times the value derived from
observations, a huge discrepancy known as ``cosmological constant problem''%
\cite{Weinberg}, whose most simple solution is to assume a cancelation
between positive and negative terms of the vacuum energy. However, as
Weinberg argued\cite{Weinberg}, it seems difficult to believe that the
cancelation is not exact but it is fine tuned to reduce the disagreement by
124 orders of magnitude. The problem might be an argument against the
reality of the vacuum fields. Nevertheless the difficulty with the fine
tuning is solved in a natural way if we assume that the \textit{mean vacuum
energy and pressure }are exactly balanced but the cosmological constant
derives from the vacuum \textit{fluctuations}. In fact, at a difference with
Newtonian gravity, a linear theory where a fluctuating mass with zero mean
would produce a fluctuating field also with zero mean, general relativity is
non-linear and therefore a fluctuating energy density with zero mean may
give rise to a gravitational field (space-time curvature) with mean
different from zero\cite{Santos11}.

Actually a possible cancelation of the mean vacuum energy might exist
between the positive contribution of bose fields and the negative one of the
fermi fields. The positive contribution of the best known bose field, the
electromagnetic radiation, was exhibited in eq.$\left( \ref{00}\right) ,$
now I illustrate the negative contribtion of fermi fields with the example
of the positron-electron field. In this case the free Hamiltonian may be
written\cite{Sakurai} 
\begin{equation}
H=\sum_{\mathbf{p,}s}E_{\mathbf{p}}\left( b_{\mathbf{p}}^{\left( s\right)
\dagger }b_{\mathbf{p}}^{\left( s\right) }-d_{-\mathbf{p}}^{\left( s\right)
}d_{\mathbf{-p}}^{\left( s\right) \dagger }\right) =\sum_{\mathbf{p,}s}E_{%
\mathbf{p}}\left( b_{\mathbf{p}}^{\left( s\right) \dagger }b_{\mathbf{p}%
}^{\left( s\right) }+d_{\mathbf{p}}^{\left( s\right) \dagger }d_{\mathbf{p}%
}^{\left( s\right) }-1\right) ,\smallskip \smallskip \smallskip
\label{Hdirac}
\end{equation}
where $b$ ($d$) are electron (positron) creation or annihilation operators, $%
\mathbf{p}$ is the momentum, $s$ the spin projection and $E=\sqrt{m^{2}+%
\mathbf{p}^{2}}$ is a (positive) energy. The latter eq.$\left( \ref{Hdirac}%
\right) $ consists of terms in normal ordering, that do not contribute to
the vacuum expectation value, plus a term that gives a \textit{negative}
energy contribution, so that the vacuum expectation of the Hamiltonian is $%
-\sum_{\mathbf{p},s}E_{\mathbf{p}}$, a negative divergent quantity. The
results eqs.$\left( \ref{00}\right) $ and $\left( \ref{Hdirac}\right) $ are
illustrative of the possible cancelation of positive (bose fields) and
negative (fermi fields) contributions to the quantum vacuum energy. Of
course the assumed cancelation should involve all possible vacuum fields and
their interactions. Similarly the positive and negative pressures of bose
and fermi vacuum fields also should cancel each other.

The existence of positive and negative energy density and pressure
contributions of the quantum vacuum fields suggests that the vacuum could be
gravitationally polarized, in a way similar to the electric polarization
that gives a contribution to the Lamb shift in heavy atoms. The electric
force, being atractive between opposite charges, produces a screening that
diminish the effective charge of either a ion in water or a nucleus in
vacuum. However gravity is always attractive thus enhancing the field near a
mass, whence the said polarization would produce an \textit{additional}
gravitational field in the neighbourhood of big masses like galaxies or
clusters. A simplified model has been proposed suggesting that this
mechanism may be an alternative to the assumed dark matter\cite{Santos18}.

In summary there are strong arguments for the reality of the quantum vacuum
fields possessing energy and pressure. Firstly the vacuum fields appear
naturally in quantization and should not be removed artificially. Secondly
the action of the vacuum fields in laboratory experiments may be weak,
usually not observable, due to an almost complete cancelation between
radiation traveling in opposite directions, but it may be measured in some
delicate breakings of balance like the Casimir effect. In astrophysics and
cosmology the alleged huge energy and pressure of the vacuum fields may not
hold, because a cancelation could exists between positive and negative
contributions.

\section{Entangled photon experiments in the Hilbert-space formalism of
quantum optics}

Spontaneous parametric down-conversion (SPDC), where a crystal having
nonlinear electric susceptibility is pumped by a laser, is the common method
to produce entangled photon pairs. Amongst the radiation emitted from the
opposite side of the crystal two beams are selected, named ``signal'' and
``idler'', with appropriate apertures and lens system. These beams are
currently interpreted as consisting of a flow of entangled photon pairs
derived from the photons of the laser, with energy $
\rlap{\protect\rule[1.1ex]{.325em}{.1ex}}h%
\omega _{P}$ each, that are split by the coupling with the crystal, giving
rise to two photons with energies $
\rlap{\protect\rule[1.1ex]{.325em}{.1ex}}h%
\omega _{s}$ and $
\rlap{\protect\rule[1.1ex]{.325em}{.1ex}}h%
\omega _{i},$ $\omega _{s}$+$\omega _{i}=\omega _{P}$, this equality
interpreted as energy conservation. In contrast the photon momenta are not
conserved because the crystal takes a part of the momentum. Indeed the
emerging signal and idler photons travel in different directions so that
their joint momentum cannot be zero.

In order to study the phenomenon within quantum optics it is standard to
consider that the laser and two ``vacuum beams'' enter the crystal, interact
with the laser and emerge as ``signal and idler'' beams\cite{C1}, \cite
{Milonni}. In a simplified two-modes treatment the incoming and emerging
beams are represented by two radiation modes with associated field operators 
\begin{equation}
incoming:\hat{a}_{s},\hat{a}_{i};outgoing:\hat{a}_{s}+D\hat{a}_{i}^{\dagger
},\hat{a}_{i}+D\hat{a}_{s}^{\dagger },  \label{3.1}
\end{equation}
where $D$ is a small coupling parameter, i. e. $\left| D\right| $ $<<1$. The
incoming modes, corresponding to vacuum fields, are ``inactive'' and so
represented by annihilation operators whilst the outgoing modes have
inactive parts plus actual photons, i. e. these with associated creation
operators.

The use of two modes as in eq.$\left( \ref{3.1}\right) $ provides a bad
representation of the physics. In fact a physical beam corresponds to a
superposition of the amplitudes, $\hat{a}_{\mathbf{k}}^{\dagger },$ of many
modes with frequencies and wavevectors close to $\omega _{s}$ and $\mathbf{k}%
_{s},$ respectively. For instance we may represent the positive frequency
part of an idler beam created in the crystal, to first order in $D$, as
follows 
\begin{equation}
\hat{E}_{i}^{\left( +\right) }\left( \mathbf{r},t\right) =D\int f_{i}\left( 
\mathbf{k}\right) d^{3}\mathbf{k}\hat{a}_{\mathbf{k}}^{\dagger }\exp \left[
i\left( \mathbf{k}-\mathbf{k}_{s}\right) \mathbf{\cdot r}-i\left( \omega
-\omega _{s}\right) t\right] +\hat{E}_{ZPF}^{\left( +\right) },  \label{55}
\end{equation}
where $\omega =\omega \left( \mathbf{k}\right) $ and $f_{i}\left( \mathbf{k}%
\right) $ is an appropriate function, with domain in a region of $\mathbf{k}$
near $\mathbf{k}_{s}.$ The field $\hat{E}_{ZPF}^{\left( +\right) }$ is the
sum of amplitudes of all vacuum modes, including the one represented by $%
\hat{a}_{s}$ in eq.$\left( \ref{3.1}\right) .$ Nevertheless the two-mode
approximation is generally good enough for calculations. An exception will
appear in section 5, when we evaluate eq.$\left( \ref{s16}\right) .$ In the
rest of the calculations we will ignore the space-time phase factor present
in eq.$\left( \ref{55}\right) $.

With appropriate devices the signal and idler beams may be combined giving
rise to two beams with photons entangled in polarization. These beams travel
(usually a long path) until Alice and Bob respectively. Alice possesses a
polarization analyzer and a detector in front of it, and similarly Bob. Thus
the beam fields arriving at these two detectors may be represented, in our
two modes approximation, by the operators 
\begin{eqnarray}
Alice &:&\hat{E}_{A}^{+}=\hat{a}_{s}\cos \theta +i\hat{a}_{i}\sin \theta +D[%
\hat{a}_{i}^{\dagger }\cos \theta +i\hat{a}_{s}^{\dagger }\sin \theta ], 
\nonumber \\
Bob &:&\hat{E}_{B}^{+}=-i\hat{a}_{i}\cos \phi +\hat{a}_{s}\sin \phi +D[-i%
\hat{a}_{s}^{\dagger }\cos \phi +\hat{a}_{i}^{\dagger }\sin \phi ],
\label{3.2}
\end{eqnarray}
where $\theta $ and $\phi $ are the polarizer\'{}s angles of Alice and Bob
respectively. The Hermitean conjugate of these field operators will be
labelled $\hat{E}_{A}^{-}\equiv \left( \hat{E}_{A}^{+}\right) ^{\dagger },$ $%
\hat{E}_{B}^{-}\equiv \left( \hat{E}_{B}^{+}\right) ^{\dagger }.$

From eqs.$\left( \ref{3.2}\right) $ it is straightforward to get the quantum
predictions for the experiment using the standard Hilbert-space formalism
(HS in the following). Alice single detection rate is proportional to the
following vacuum expectation 
\begin{equation}
R_{A}=\langle 0\mid \hat{E}_{A}^{-}\hat{E}_{A}^{+}\mid 0\rangle =\left|
D\right| ^{2}\langle 0\mid \hat{a}_{i}\hat{a}_{i}^{\dagger }\cos ^{2}\theta +%
\hat{a}_{s}\hat{a}_{s}^{\dagger }\sin ^{2}\theta \mid 0\rangle =\left|
D\right| ^{2},  \label{3.3}
\end{equation}
where I have neglected terms with creation (annihilation) operators
appearing on the left (right). A similar result may be obtained for the
single detection rate of Bob, that is 
\begin{equation}
R_{B}=\langle 0\mid \hat{E}_{B}^{-}\hat{E}_{B}^{+}\mid 0\rangle =\left|
D\right| ^{2}.  \label{3.4}
\end{equation}

The coincidence rate is obtained via the vacuum expectation value of the
product of four field operators in normal order. In our case we have two
terms, that is 
\begin{equation}
R_{AB}=\frac{1}{2}\langle 0\mid \hat{E}_{A}^{-}\hat{E}_{B}^{-}\hat{E}_{B}^{+}%
\hat{E}_{A}^{+}\mid 0\rangle +\frac{1}{2}\langle 0\mid \hat{E}_{B}^{-}\hat{E}%
_{A}^{-}\hat{E}_{A}^{+}\hat{E}_{B}^{+}\mid 0\rangle ,  \label{3.5}
\end{equation}
which would be equal to each other if $\hat{E}_{A}^{+}$ and $\hat{E}_{B}^{+}$
commuted. The former expectation may be evaluated to order $\left| D\right|
^{2}$ as follows 
\begin{eqnarray}
\langle 0 &\mid &\hat{E}_{A}^{-}\hat{E}_{B}^{-}\hat{E}_{B}^{+}\hat{E}%
_{A}^{+}\mid 0\rangle =\langle 0\mid \hat{E}_{A1}^{-}\hat{E}_{B}^{-}\hat{E}%
_{B}^{+}\hat{E}_{A1}^{+}\mid 0\rangle  \nonumber \\
&=&\langle 0\mid \hat{E}_{A1}^{-}\hat{E}_{B0}^{-}\hat{E}_{B0}^{+}\hat{E}%
_{A1}^{+}\mid 0\rangle =\left| \langle 0\mid \hat{E}_{B0}^{+}\hat{E}%
_{A1}^{+}\mid 0\rangle \right| ^{2},  \label{3.6}
\end{eqnarray}
where $\hat{E}_{A1}^{+}$ is the part of order $\left| D\right| $ of $\hat{E}%
_{A}^{+}$ and $\hat{E}_{B1}^{+}$ the part of order $\left| D\right| $ of $%
\hat{E}_{B}^{+}$, see eq.$\left( \ref{3.2}\right) ,$ $\hat{E}_{A0}^{+}$ and $%
\hat{E}_{B0}^{+}$ being the parts of order zero respectively. In the former
equality of eq.$\left( \ref{3.6}\right) $ we have removed creation operators
on the left and annihilation operators on the right, in the second we have
removed terms of order $\left| D\right| ^{4}.$ The latter term of eq.$\left( 
\ref{3.5}\right) $ gives a similar result, with $A(B)$ substituted for $B(A)$%
. Then the coincidence detection rate becomes 
\begin{equation}
R_{AB}=\frac{1}{2}\left| \langle 0\mid \hat{E}_{B0}^{+}\hat{E}_{A1}^{+}\mid
0\rangle \right| ^{2}+\frac{1}{2}\left| \langle 0\mid \hat{E}_{A0}^{+}\hat{E}%
_{B1}^{+}\mid 0\rangle \right| ^{2}.  \label{3.8}
\end{equation}

Hence, taking eqs.$\left( \ref{3.2}\right) $ into account we have 
\[
\left| \langle 0\mid \hat{E}_{B0}^{+}\hat{E}_{A1}^{+}\mid 0\rangle \right|
^{2}=\left| D\right| ^{2}\left| \sin \phi \sin \theta +\cos \phi \cos \theta
\right| ^{2}=\left| D\right| ^{2}\cos ^{2}(\theta -\phi ). 
\]
The latter term of eq.$\left( \ref{3.8}\right) $ leads to a similar
contribution whence we get 
\begin{equation}
R_{AB}=\left| D\right| ^{2}\cos ^{2}(\theta -\phi ).  \label{3.7}
\end{equation}
The results eqs.$\left( \ref{3.3}\right) ,\left( \ref{3.4}\right) $ and $%
\left( \ref{3.7}\right) $ have the form of eq.$\left( \ref{pred}\right) $
and therefore the quantum predictions violate a Bell inequality.

\section{The experiments in the Weyl-Wigner formalism}

\subsection{The formalism in quantum optics}

In the following I shall shortly review the treatment within the Weyl-Wigner
(WW) formalism of the polarization correlation experiment. The WW formalism
was developped for non-relativistic quantum mechanics, where the basic
observables involved are positions, $\hat{x}_{j},$ and momenta, $\hat{p}%
_{j}, $ of the particles\cite{Weyl}, \cite{Wigner}, \cite{Zachos}. It may be
trivially extended to quantum optics provided we interpret $\hat{x}_{j}$ and 
$\hat{p}_{j}$ to be the sum and the difference of the creation, $\hat{a}%
_{j}^{\dagger },$ and annihilation, $\hat{a}_{j},$ operators of the $j$
normal mode of the radiation. That is

\begin{eqnarray}
\hat{x}_{j} &\equiv &\frac{c}{\sqrt{2}\omega _{j}}\left( \hat{a}_{j}+\hat{a}%
_{j}^{\dagger }\right) ,\hat{p}_{j}\equiv \frac{i 
\rlap{\protect\rule[1.1ex]{.325em}{.1ex}}h%
\omega _{j}}{\sqrt{2}c}\left( \hat{a}_{j}-\hat{a}_{j}^{\dagger }\right) 
\nonumber \\
&\Rightarrow &\hat{a}_{j}=\frac{1}{\sqrt{2}}\left( \frac{\omega _{j}}{c}\hat{%
x}_{j}+\frac{ic}{
\rlap{\protect\rule[1.1ex]{.325em}{.1ex}}h%
\omega }\hat{p}_{j}\right) ,\hat{a}_{j}^{\dagger }=\frac{1}{\sqrt{2}}\left( 
\frac{\omega _{j}}{c}\hat{x}_{j}-\frac{ic}{
\rlap{\protect\rule[1.1ex]{.325em}{.1ex}}h%
\omega _{j}}\hat{p}_{j}\right) .  \label{4.1}
\end{eqnarray}
Here $
\rlap{\protect\rule[1.1ex]{.325em}{.1ex}}h%
$ is Planck constant, $c$ the velocity of light and $\omega _{j}$ the
frequency of the normal mode. In the following I will use units $
\rlap{\protect\rule[1.1ex]{.325em}{.1ex}}h%
=c=1$. For the sake of clarity I will represent the field operators on a
Hilbert space with a `hat' as in the previous section, e. g. $\hat{a}_{j},%
\hat{a}_{j}^{\dagger }$, but remove the `hat' for the amplitudes in the WW
formalism, e. g. $a_{j},a_{j}^{*}.$

The connection with the Hilbert-space (HS) formalism is made via the Weyl
transform as follows. For any trace class operator $\hat{M}$ of the former
we define its Weyl transform to be a function of the field operators $%
\left\{ \hat{a}_{j},\hat{a}_{j}^{\dagger }\right\} $, that is 
\begin{eqnarray}
W_{\hat{M}} &=&\frac{1}{(2\pi ^{2})^{n}}\prod_{j=1}^{n}\int_{-\infty
}^{\infty }d\lambda _{j}\int_{-\infty }^{\infty }d\mu _{j}\exp \left[
-2i\lambda _{j}\mathrm{Re}a_{j}-2i\mu _{j}\mathrm{Im}a_{j}\right]  \nonumber
\\
&&\times Tr\left\{ \hat{M}\exp \left[ i\lambda _{j}\left( \hat{a}_{j}+\hat{a}%
_{j}^{\dagger }\right) +i\mu _{j}\left( \hat{a}_{j}-\hat{a}_{j}^{\dagger
}\right) \right] \right\} .  \label{4.2}
\end{eqnarray}
The transform is \textit{invertible} that is 
\begin{eqnarray*}
\hat{M} &=&\frac{1}{(2\pi ^{2})^{2n}}\prod_{j=1}^{n}\int_{-\infty }^{\infty
}d\lambda _{j}\int_{-\infty }^{\infty }d\mu _{j}\exp \left[ i\lambda
_{j}\left( \hat{a}_{j}+\hat{a}_{j}^{\dagger }\right) +i\mu _{j}\left( \hat{a}%
_{j}-\hat{a}_{j}^{\dagger }\right) \right] \\
&&\times \prod_{j=1}^{n}\int_{-\infty }^{\infty }d\mathrm{Re}%
a_{j}\int_{-\infty }^{\infty }d\mathrm{Im}a_{j}W_{\hat{M}}\left\{
a_{j},a_{j}^{*}\right\} \exp \left[ -2i\lambda _{j}\mathrm{Re}a_{j}-2i\mu
_{j}\mathrm{Im}a_{j}\right] .
\end{eqnarray*}
The transform is\textit{\ linear}, that is if $f$ is the transform of $\hat{f%
}$ and $g$ the transform of $\hat{g}$, then the transform of $\hat{f}$ +$%
\hat{g}$ is $f+g$.

It is standard wisdom that the WW formalism is unable to provide any
intuitive picture of the quantum phenomena. The reason is that the Wigner
function, that would represent the quantum states of the HS formalism, is
not positive definite in general and therefore cannot be interpreted as a
probability distribution (of positions and momenta in quantum mechanics, or
field amplitudes in quantum optics). However we shall see that in quantum
optics the formalism in the ``Heisenberg pcture'' allows the interpretation
of the experiments using the Wigner function only for the vacuum state, that
is positive definite.

The use of the WW formalism in quantum optics has the following features in
comparison with the HS formalism:

1. It is just quantum optics, therefore the predictions for experiments are
the same.

2. The calculations using the WW formalism are usually no more involved than
the corresponding ones in Hilbert space. However the latter allows
significant shortcuts in some cases via a clever use of the noncommutative
algebra of operators in HS.

3. The formalism suggests a physical picture in terms of random variables
and stochastic processes. In particular the counterparts of creation and
annihilation operators look like random amplitudes in a complex
representation of radiation.

However I\ shall stress that the physical picture is possible only if we
renounce to get an interpretation in terms of random variables for all
alleged states and observables of the standard HS formalism, but we attempt
just to \textit{interpret (get a physical picture of) actual experiments, }%
either performed or possible. An example is the interpretation offered in
section 5 for test of Bell inequalities without photons, although I will use
sometimes the common language speaking for instance about ``entangled
photons'' in order to show the connection with the HS formalism. In our
interpretation of the experiments the particle behaviour of the quantum
electromagnetic field does not appear at all, photocounts being just rapid
changes in the detectors due to the interaction with the radiation (wave)
field.

\subsection{Properties}

All properties of the WW formalism in particle systems may be translated to
quantum optics via eqs.$\left( \ref{4.1}\right) .$ The transform eq.$\left( 
\ref{4.2}\right) $ allows getting a function of (c-number) amplitudes for
any trace-class operator ( e. g. any function of the creation and
annihilation operators of `photons'). In particular we may get the (Wigner)
function corresponding to any quantum state of the HS formalism. For
instance the vacuum state, represented by the density matrix $\left|
0\rangle \langle 0\right| ,$ is associated to the following Wigner function 
\begin{equation}
W_{0}=\prod_{j}\frac{2}{\pi }\exp \left( -2\left| a_{j}\right| ^{2}\right) .
\label{1}
\end{equation}
Hence the suggested picture that the quantum vacuum of the electromagnetic
field (the zeropoint field\textit{, ZPF}) consists of stochastic fields with
a probability distribution independent for every mode, having a Gaussian
distribution with mean energy $\frac{1}{2}
\rlap{\protect\rule[1.1ex]{.325em}{.1ex}}h%
\omega $ per mode. That interpretation will be studied in the next section.

Similarly there are functions associated to the observables. For instance
the following Weyl transforms are obtained 
\begin{eqnarray}
\hat{a}_{j} &\leftrightarrow &a_{j},\hat{a}_{j}^{\dagger }\leftrightarrow
a_{j}^{*},\frac{1}{2}\left( \hat{a}_{j}^{\dagger }\hat{a}_{j}+\hat{a}_{j}%
\hat{a}_{j}^{\dagger }\right) \leftrightarrow a_{j}a_{j}^{*}=\left|
a_{j}\right| ^{2},  \nonumber \\
\hat{a}_{j}^{\dagger }\hat{a}_{j} &=&\frac{1}{2}\left( \hat{a}_{j}^{\dagger }%
\hat{a}_{j}+\hat{a}_{j}\hat{a}_{j}^{\dagger }\right) +\frac{1}{2}\left( \hat{%
a}_{j}^{\dagger }\hat{a}_{j}-\hat{a}_{j}\hat{a}_{j}^{\dagger }\right)
\leftrightarrow \left| a_{j}\right| ^{2}-\frac{1}{2},  \nonumber \\
\left( \hat{a}_{j}^{\dagger }+\hat{a}_{j}\right) ^{n} &\leftrightarrow
&\left( a_{j}+a_{j}^{*}\right) ^{n},\left( \hat{a}_{j}^{\dagger }-\hat{a}%
_{j}\right) ^{n}\leftrightarrow \left( a_{j}-a_{j}^{*}\right) ^{n},n\text{
an integer.}  \label{2}
\end{eqnarray}
I stress that the quantities $a_{j}$ and $a_{j}^{*}$ are c-numbers and
therefore they commute with each other. The former eqs.$\left( \ref{2}%
\right) $ mean that in expressions \textit{linear in creation and/or
annihilation operator} the Weyl transform just implies `\textit{removing the
hats}'. However this is not the case in nonlinear expressions in general. In
fact from the latter two eqs.$\left( \ref{2}\right) $ plus the linear
property it follows that for a product in the WW formalism the HS
counterpart is 
\begin{equation}
a_{j}^{k}a_{j}^{*^{l}}\leftrightarrow (\hat{a}_{j}^{k}\hat{a}_{j}^{\dagger
l})_{sym},  \label{2b}
\end{equation}
where the subindex $sym$ means writing the product in all possible orderings
and dividing by the number of terms. Hence the WW field amplitudes
corresponding to products of field operators may be obtained putting the
operators in symmetrical order via the commutation relations.

\textit{Expectation values} may be calculated in the WW formalism as
follows. In the HS formalism they read $Tr(\hat{\rho}\hat{M})$, or in
particular $\langle \psi \mid \hat{M}\mid \psi \rangle ,$ whence the
translation to the WW formalism is obtained taking into account that the
trace of the product of two operators becomes 
\begin{equation}
Tr(\hat{\rho}\hat{M})=\int W_{\hat{\rho}}\left\{ \hat{a}_{j},\hat{a}%
_{j}^{\dagger }\right\} W_{\hat{M}}\left\{ \hat{a}_{j},\hat{a}_{j}^{\dagger
}\right\} \prod_{j}d\mathrm{Re}a_{j}d\mathrm{Im}a_{j}.  \label{3}
\end{equation}
That integral is the WW counterpart of the trace operation in the HS
formalism. Particular instances are the following expectations that will be
of interest later on 
\begin{eqnarray}
\left\langle \left| a_{j}\right| ^{2}\right\rangle &\equiv &\int d\Gamma
W_{0}\left| a_{j}\right| ^{2}=\frac{1}{2},\left\langle
a_{j}^{n}a_{j}^{*m}\right\rangle =0\text{ if }n\neq m.  \nonumber \\
\left\langle 0\left| \hat{a}_{j}^{\dagger }\hat{a}_{j}\right| 0\right\rangle
&=&\int d\Gamma (a_{j}^{*}a_{j}-\frac{1}{2})W_{0}=0,  \nonumber \\
\left\langle 0\left| \hat{a}_{j}\hat{a}_{j}^{\dagger }\right| 0\right\rangle
&=&\int d\Gamma (\left| a_{j}\right| ^{2}+\frac{1}{2})W_{0}=2\left\langle
\left| a_{j}\right| ^{2}\right\rangle =1,  \label{3a} \\
\left\langle \left| a_{j}\right| ^{4}\right\rangle &=&1/2,\left\langle
\left| a_{j}\right| ^{n}\left| a_{k}\right| ^{m}\right\rangle =\left\langle
\left| a_{j}\right| ^{n}\right\rangle \left\langle \left| a_{k}\right|
^{m}\right\rangle \text{ if }j\neq k.  \nonumber
\end{eqnarray}
where $W_{0}$ is the Wigner function of the vacuum, eq.$\left( \ref{1}%
\right) $. This means that in the WW formalism the field amplitude $a_{j}$
(coming from the vacuum) behaves like a complex random variable with
Gaussian distribution and mean square modulus $\left\langle \left|
a_{j}\right| ^{2}\right\rangle =1/2.$ I point out that the integral for any
mode not entering in the function $W_{\hat{\rho}}\left( \left\{
a_{j},a_{j}^{*}\right\} \right) $ gives unity in the integral eq.$\left( \ref
{3}\right) $ due to the normalization of the Wigner function eq.$\left( \ref
{1}\right) $. An important consequence of eq.$\left( \ref{3a}\right) $ is
that normal (antinormal) ordering of one creation and one annihilation
operators in the Hilbert space formalism becomes subtraction (addition) of
1/2 in the WW formalism. That is the normal ordering rule is equivalent to
subtracting the vacuum contribution, as was commented in section 2.1.

\subsection{ Entangled photon pairs in the WW formalism}

All quantum optical phenomena that may be analyzed using the HS formulation
of quantum optics may be also studied with the WW formalism. In fact, it is
enough to translate the equations to the new formalism by means of the Weyl
transform. Here I will apply the WW formalism to the description of the
polarization correlation of entangled photon pairs produced via spontaneous
parametric down-conversion (SPDC). I will start with the fields that are the
WW counterparts of the HS field operators eqs.$\left( \ref{3.2}\right) $,
that I will write in terms of two partial field amplitudes for later
convenience, that is 
\begin{eqnarray}
E_{A}^{+} &=&E_{A0}^{+}+E_{A1}^{+},E_{B}^{+}=E_{B0}^{+}+E_{B1}^{+}, 
\nonumber \\
E_{A0}^{+} &=&a_{s}\cos \theta +ia_{i}\sin \theta ,E_{A1}^{+}=D\left[
a_{i}^{*}\cos \theta +ia_{s}^{*}\sin \theta \right] ,  \nonumber \\
E_{B0}^{+} &=&-ia_{s}\sin \phi +a_{i}\cos \phi ,E_{B1}^{+}=D\left[
-ia_{i}^{*}\sin \phi +a_{s}^{*}\cos \phi \right] .  \label{b3}
\end{eqnarray}

Now we are in a position to derive the WW detection rules via a Weyl
transform of the stated rules in the HS formalism. The following transforms
may be easily derived from eqs.$\left( \ref{3a}\right) $ 
\begin{eqnarray}
\langle 0 &\mid &\hat{a}_{i}\hat{a}_{i}^{\dagger }\mid 0\rangle \rightarrow
\langle \left| a_{i}\right| ^{2}+\frac{1}{2}\rangle =2\langle \left|
a_{i}\right| ^{2}\rangle ,  \nonumber \\
\langle 0 &\mid &\hat{a}_{i}^{\dagger }\hat{a}_{i}\mid 0\rangle \rightarrow
\langle \left| a_{i}\right| ^{2}-\frac{1}{2}\rangle =0.  \label{e}
\end{eqnarray}
Hence it may be realized that the HS expectations with the field operators
in antinormal order, as in eqs.$\left( \ref{3.3}\right) $ or $\left( \ref
{3.8}\right) ,$ a factor 2 should be included for every pair of field
operators. Therefore eq.$\left( \ref{3.3}\right) $ leads to 
\begin{equation}
R_{A}=\langle 0\mid \hat{E}_{A1}^{-}\hat{E}_{A1}^{+}\mid 0\rangle
\rightarrow R_{A}=2\langle E_{A1}^{-}E_{A1}^{+}\rangle ,R_{B}=2\langle
E_{B1}^{-}E_{B1}^{+}\rangle .  \label{e1}
\end{equation}
With analogous arguments the HS rule for the coincidence rate, eq.$\left( 
\ref{3.6}\right) ,$ should be multiplied times 4, giving the WW rule 
\begin{eqnarray}
R_{AB} &=&2\left| \langle E_{A0}^{+}E_{B1}^{+}\rangle \right| ^{2}+2\left|
\langle E_{B0}^{+}E_{A1}^{+}\rangle \right| ^{2}  \label{e2} \\
&=&2\left| \langle E_{A0}^{+}E_{B}^{+}\rangle \right| ^{2}+2\left| \langle
E_{B0}^{+}E_{A}^{+}\rangle \right| ^{2},
\end{eqnarray}
where the latter equality takes into account the identity 
\[
\langle E_{A0}^{+}E_{B}^{+}\rangle =\langle
E_{A0}^{+}(E_{B0}^{+}+E_{B1}^{+})\rangle =\langle
E_{A0}^{+}E_{B0}^{+}\rangle +\langle E_{A0}^{+}E_{B1}^{+}\rangle , 
\]
and $\langle E_{A0}^{+}E_{B0}^{+}\rangle =0$ that may be derived from the
second eq.$\left( \ref{3a}\right) $ with $n=2,m=0$.

Eq.$\left( \ref{e2}\right) $ provides the desired coincidence detection
rule, which is rather simple written in terms of field amplitudes arriving
at Alice and Bob respectively. By construction it is obvious that, using eqs.%
$\left( \ref{e1}\right) $ and $\left( \ref{e2}\right) ,$ the WW formalism
will give the same predictions as the standard quantum HS formalism for all
experiments involving entangled photon pairs produced via SPDC.

For the realistic interpretation to be given in the next section it is
interesting to write the detection rules eqs.$\left( \ref{e1}\right) $ and $%
\left( \ref{e2}\right) $ in terms of field intensities rather than
amplitudes. To do that we may define intensities as follows

\begin{eqnarray}
I_{A0}
&=&E_{A0}^{+}E_{A0}^{-},I_{A1}=E_{A0}^{+}E_{A1}^{-}+E_{A1}^{+}E_{A0}^{-},I_{A2}=E_{A1}^{+}E_{A1}^{-},
\nonumber \\
I_{B0}
&=&E_{B0}^{+}E_{B0}^{-},I_{B1}=E_{B0}^{+}E_{B1}^{-}+E_{B1}^{+}E_{B0}^{-},I_{B2}=E_{B1}^{+}E_{B1}^{-},
\nonumber \\
I_{A}
&=&E_{A}^{+}E_{A}^{-}=I_{A0}+I_{A1}+I_{A2},I_{B}=E_{B}^{+}E_{B}^{-}=I_{B0}+I_{B1}+I_{B2},
\label{b5}
\end{eqnarray}
although $I_{A1}$ and $I_{B1}$ are not actual intensities, in particular
they are not positive definite. I point out that $I_{A0},I_{A1}$ and $I_{A2}$
are of order $1,\left| D\right| $ and $\left| D\right| ^{2},$ respectively,
in the small parameter $\left| D\right| <<1.$ Eq.$\left( \ref{e1}\right) $
may trivially obtained in terms of intensities taking eqs.$\left( \ref{b5}%
\right) $ into account. We get 
\begin{equation}
R_{A}=2\langle I_{A}\rangle -2\langle I_{A0}\rangle =2\langle I_{A2}\rangle
,R_{B}=2\langle I_{B}\rangle -2\langle I_{B0}\rangle =2\langle I_{B2}\rangle
,  \label{e11}
\end{equation}
the terms $I_{A1}$ and $I_{B1}$ not contributing, as may be realized.
Writing the coincidence detection rate eq.$\left( \ref{e2}\right) $ in terms
of intensities is more involved and will be postponed to the next section.

We conclude that the predictions for the experiments are the same either in
the HS or in the WW formalism provided in the latter we use for the single
rates either eq.$\left( \ref{e1}\right) $ or eq.$\left( \ref{e11}\right) $
and for the coincidence rates eq.$\left( \ref{e2}\right) .$

\section{A realistic local model of the experiments}

\subsection{Model of photodetector}

The WW formalism suggests a picture of the quantum optical phenomena. The
picture for experiments involving `entangled photon pairs' provides a local
realistic interpretation in terms of random variables and stochastic
processes. In the following I present the main ideas of this stochastic
interpretation. It rests upon several assumptions as follows.

The fundamental hypothesis is that the electromagnetic vacuum field is a
real stochastic field (the zeropoint field, ZPF) as commented in section 2.
If expanded in normal modes the ZPF has a (positive) probability
distribution of the amplitudes given by eq.$\left( \ref{1}\right) .$
Therefore we assume that all bodies are immersed in ZPF, charged particles
absorbing and emiting radiation continuously, in some cases reaching a
dynamical equilibrium with the ZPF (that would correspond to the ground
state in the HS formalism\cite{dice}).

According to that assumption any photodetector would be under the action of
an extremely strong radiation, infinite if no cut-off existed. Thus how
might we explain that detectors are not activated by the vacuum radiation?
Firstly the strong vacuum field is effectively reduced to a weak level if 
\textit{we assume that only radiation within some small frequency interval
is able to activate a photodetector}, that is the interval of sensibility $%
\left( \omega _{1},\omega _{2}\right) $. However the problem is not yet
solved because signals involved in experiments have typical intensities of
order the vacuum radiation in the said frequency interval so that the
detector would be unable to distinguish a signal from the ZPF noise. Our
proposed solution is to assume that \textit{a detector may be activated only
when the net Poynting vector (i. e. the directional energy flux) of the
incoming radiation is different from zero, including both signal and vacuum
fields}. More specifically I will model a\textit{\ detector as possessing an
active area, the probability of a photocount being proportional to the net
radiant energy flux crossing that area from the front side during the
activation time,\ the probability being zero if the net flux crosses the
area in the reverse direction during that time interval. }The need of some
finite detection time is a known fact in experiments.

These assumptions allow to understand qualitatively why the signals, but not
the vacuum fields, activate detectors. Indeed the ZPF arriving at any point
(in particular the detector) would be usually isotropic on the average,
whence its associated mean Poynting vector would be nil, therefore only the
(directional) signal radiation might produce photocounts. A problem remains
because the vacuum fields are fluctuating so that the net Poynting vector
also fluctuates, and it may point in the wrong direction even in the
presence of signals. However a net flux in the wrong direction would be
unlikely\textit{\ if the activation time of the detectors is large enough}
because this would effectively average the vacuum fluctuations, washing
their effect.

Our aim is to achieve a realistic local interpretation of the experiments
measuring polarization correlation of entangled photon pairs, that we
studied with the WW formalism in the previous section. Thus I will consider
two vacuum beams entering the nonlinear crystal, where they give rise to a
``signal'' and an ``idler'' beams. After crossing several appropriate
devices they produce fields that arrive at the Alice and Bob detectors. I
will not attempt a detailed model that should involve many modes in order to
represent the signals as narrow beams (see eq.$\left( \ref{55}\right) )$.

In agreement with our previous hypotheses a photocount should derive from
the net energy flux crossing the active photocounter surface. Thus I will
assume that the detection probabilities per time window, $T$, that are
proportional to the single $R_{A},R_{B}$ and coincidence $R_{AB}$ detection
rates, will be 
\begin{eqnarray}
R_{A} &=&\left\langle \left[ M_{A}\right] _{+}\right\rangle
,R_{B}=\left\langle \left[ M_{A}\right] _{+}\right\rangle
,R_{AB}=\left\langle \left[ M_{A}\right] _{+}\left[ M_{B}\right]
_{+}\right\rangle ,  \nonumber \\
M_{A} &\equiv &T^{-1}\int_{0}^{T}\vec{n}_{A}\cdot \vec{I}_{total}^{A}\left( 
\mathbf{r}_{A}\mathbf{,}t\right) dt,M_{B}\equiv T^{-1}\int_{0}^{T}\vec{n}%
_{B}\cdot \vec{I}_{total}^{B}\left( \mathbf{r}_{A},t\right) dt,  \label{s1}
\end{eqnarray}
where $\left[ M\right] _{+}=$ $M$ if $M>0,$ $\left[ M\right] _{+}=0$
otherwise, and $\vec{n}$ is a unit vector in the direction of the incoming
signal beams, that I assume perpendicular to the active area of the
detector. I use units such that both the intensities and the detection rates
are dimensionless, the latter because they are defined as probabilities
within a time window $T,$ this being greater than the photocounter
activation time. In eq.$\left( \ref{s1}\right) $ the positivity constraint
(i.e. putting $\left[ M_{A}\right] _{+}$ rather than $M_{A})$ is needed
because the detection probabilities must be non-negative for any particular
run of an experiment whilst the quantities $M_{A}$ and $M_{B}$ are
fluctuating and might be negative. Nevertheless the ensemble averages
involved in eq.$\left( \ref{s1}\right) $ are positive or zero and the
fluctuations will not be too relevant due to the time integration that
washes them out. Therefore I will make the approximation of ignoring the
positivity constraint in the following, substituting $M_{A}$ for $\left[
M_{A}\right] _{+}$ and $M_{B}$ for $\left[ M_{B}\right] _{+}$ .

\subsection{Realistic interpretation of entangled photons}

In order to have a realistic model of the experiments I will consider a
simplified treatement involving just two vacuum radiation modes, with
amplitudes $a_{s}$ and $a_{i},$ as in the WW calculation of the previous
section. After crossing several appropriate devices the fields will arrive
at the Alice and Bob detectors. Each one of these two fields consists of two
parts, one of order $1$ and another of order $\left| D\right| <<1,$ see eqs.$%
\left( \ref{b3}\right) .$ It may be realized that the former is what would
arrive at the detectors if there was no pumping laser and therefore no
signal. It is just a part of the ZPF, whilst the rest of the ZPF consists of
radiation not appearing in the equations of the previous section because
they were not needed in the calculations. The total ZPF should have the
property of isotropy, therefore giving nil net flux in the detector (modulo
fluctuations that may contribute to a dark rate that we shall ignore in this
paper). The terms of order $\left| D\right| $ derive from the signals
produced in the nonlinear crystal. In summary the Poynting vectors of the
radiation at the (center of the) active area of the detectors may be written 
\begin{equation}
Alice:\vec{I}_{total}^{A}\left( t\right) =\vec{I}_{ZPF}^{A}\left( t\right) +%
\vec{I}_{A}\left( t\right) ,Bob:\vec{I}_{total}^{B}\left( t\right) =\vec{I}%
_{ZPF}^{B}\left( t\right) +\vec{I}_{B}\left( t\right) .\smallskip  \label{s2}
\end{equation}
$\vec{I}_{A}$, $\vec{I}_{B},$ are due to the fields $E_{A}$, $E_{B}$, eqs.$%
\left( \ref{b3}\right) ,$ emerging from the non-linear crystal after they
are transformed by lens systems, apertures, beam splitters, etc. The
Poynting vectors $\vec{I}_{A}\left( t\right) $ and $\vec{I}_{B}\left(
t\right) $ have the direction of $\vec{n}_{A}$ and $\vec{n}_{B\text{ }}$%
respectively, see eqs.$\left( \ref{s1}\right) ,$ and their moduli would be
field intensities. Furthermore these intensities are time independent in our
two modes representation of the fields, the time dependence in actual
experiments coming from the interference of many modes, see eq.$\left( \ref
{55}\right) $. Therefore we will write 
\begin{equation}
M_{A}=I_{ZPF}^{A}+I_{A},I_{ZPF}^{A}\equiv T^{-1}\int_{0}^{T}\vec{n}_{A}\cdot 
\vec{I}_{ZPF}^{A}\left( \mathbf{r}_{A}\mathbf{,}t\right) dt  \label{s3}
\end{equation}
and similar for $M_{B}.$

In order to get the Alice single detection rate we need the average of $%
M_{A},$ that we will evaluate by comparison with the case when there is no
pumping on the nonlinear crystal. In this case $I_{A}$ becomes $I_{A0}$ and
the Poynting vector of all vacuum fields arriving at the detector of Alice,
i. e. $\vec{I}_{ZPF}^{A}\left( t\right) +\vec{I}_{A0}\left( t\right) ,$
should have nil average due to the isotropy of the total $ZPF$. And similar
for Bob. As a consequence the intensities $I_{A0}$ and $I_{B0},$ eqs.$\left( 
\ref{b5}\right) ,$ should fulfil the following equalities 
\begin{equation}
\left\langle I_{ZPF}^{A}+I_{A0}\right\rangle =\left\langle
I_{ZPF}^{B}+I_{B0}\right\rangle =0.  \label{s4}
\end{equation}
It might appear that this relation could not be true for all values of the
angles $\theta ,\phi $ eqs.$\left( \ref{b3}\right) $ because the ZPF
Poynting vectors $\hat{I}_{ZPF}^{A}$ and $\hat{I}_{ZPF}^{B}$ should not
depend on our choice of angles whilst $I_{A0}$ and $I_{B0}$ do depend.
However the positions of the polarizers do influence also the ZPF arriving
at the detectors and it is plausible that the total Poynting vector has zero
mean in any case. From eqs.$\left( \ref{s3}\right) $ and $\left( \ref{s4}%
\right) $ we may derive the single rates of Alice and Bob, that is 
\begin{equation}
R_{A}=\left\langle M_{A}\right\rangle =\left\langle I_{A}\right\rangle
-\left\langle I_{A0}\right\rangle =\frac{1}{2}\left| D\right|
^{2},P_{B}=\left\langle M_{B}\right\rangle =\frac{1}{2}\left| D\right| ^{2}.
\label{s5}
\end{equation}
The result agrees with the quantum prediction eq.$\left( \ref{e11}\right) $
except for a scale factor 2, that will be irrelevant for our purposes as
shown later on (it derives from an arbitrary proportionally constant that we
have taken as unity in eq.$\left( \ref{s1}\right) )$.

The coincidence detection rate may be got taking eqs.$\left( \ref{s1}\right) 
$ into account, that is 
\begin{equation}
R_{AB}=\left\langle M_{A}M_{B}\right\rangle =\langle \left[
I_{ZPF}^{A}+I_{A}\right] \left[ I_{ZPF}^{B}+I_{B}\right] \rangle .
\label{s6}
\end{equation}
If there was no pumping laser on the nonlinear crystal the joint detection
rate should be zero whence eq.$\left( \ref{s6}\right) $ leads to 
\begin{equation}
\left\langle \left[ I_{ZPF}^{A}+I_{A0}\right] \left[
I_{ZPF}^{B}+I_{B0}\right] \right\rangle =0.  \label{s10}
\end{equation}
From eqs.$\left( \ref{s4}\right) ,\left( \ref{s6}\right) $ and $\left( \ref
{s10}\right) $ it is possible to get $R_{AB}.$ Firstly I point out that $%
I_{ZPF}^{A}$ and $I_{ZPF}^{B}$ could not depend on whether the pumping is
switch on. More correctly, the probability distribution of the possible
values of $I_{ZPF}$ would be the same whether the pumping is on or out. Then
subtracting eq.$\left( \ref{s10}\right) $ from eq.$\left( \ref{s6}\right) $
we get 
\[
R_{AB}=\left\langle \left( I_{B}-I_{B0}\right) I_{ZPF}^{A}\right\rangle
+\left\langle \left( I_{A}-I_{A0}\right) I_{ZPF}^{B}\right\rangle
+\left\langle I_{A}I_{B}\right\rangle -\left\langle
I_{A0}I_{B0}\right\rangle . 
\]
On the average the ZPF intensity $I_{ZPF}^{A}$ is independent of whether the
pumping is on or out, as assumed above, whence $I_{ZPF}^{A}$ and $\left(
I_{B}-I_{B0}\right) $ are uncorrelated. Similarly for $I_{ZPF}^{B}$ and $%
\left( I_{A}-I_{A0}\right) .$ Hence we may write 
\begin{eqnarray}
R_{AB} &=&\left\langle I_{A}I_{B}\right\rangle -\left\langle
I_{A0}I_{B0}\right\rangle +\left\langle I_{B}-I_{B0}\right\rangle
\left\langle I_{ZPF}^{A}\right\rangle +\left\langle
I_{A}-I_{A0}\right\rangle \left\langle I_{ZPF}^{B}\right\rangle  \nonumber \\
&=&\left\langle I_{A}I_{B}\right\rangle -\left\langle
I_{A0}I_{B0}\right\rangle -\left\langle I_{A2}\right\rangle \left\langle
I_{B0}\right\rangle -\left\langle I_{B2}\right\rangle \left\langle
I_{A0}\right\rangle ,  \label{s11}
\end{eqnarray}
the averages $\left\langle I_{A1}\right\rangle $ and $\left\langle
I_{B1}\right\rangle $ being nil. The latter inequality follows taking eq.$%
\left( \ref{s4}\right) $ into account.

Now we may write eq.$\left( \ref{s11}\right) $ in terms of the fields. For
the former term we have, taking eq.$\left( \ref{b5}\right) $ into account 
\begin{eqnarray}
\left\langle I_{A}I_{B}\right\rangle &=&\left\langle
(I_{A0}+I_{A1}+I_{A2})(I_{B0}+I_{B1}+I_{B2})\right\rangle  \nonumber \\
&=&\left\langle I_{A0}I_{B0}\right\rangle +\left\langle
I_{A0}I_{B2}\right\rangle +\left\langle I_{A2}I_{B0}\right\rangle
+\left\langle I_{A1}I_{B1}\right\rangle ,  \label{s13}
\end{eqnarray}
the terms $\left\langle I_{A0}I_{B1}\right\rangle ,\left\langle
I_{A1}I_{B0}\right\rangle ,\left\langle I_{A1}I_{B2}\right\rangle $ and $%
\left\langle I_{A2}I_{B1}\right\rangle $ not contributing as may be checked,
and the term $\left\langle I_{A1}I_{B1}\right\rangle $ contributes to order $%
\left| D\right| ^{4}$, therefore being negligible in our calculation to
order $\left| D\right| ^{2}$. The term $\left\langle
I_{A0}I_{B0}\right\rangle $ will cancel with a similar term in eq.$\left( 
\ref{s11}\right) $ and the four remaining terms of eq.$\left( \ref{s13}%
\right) $ may be written in terms of amplitudes as follows 
\begin{eqnarray*}
\left\langle I_{A0}I_{B2}\right\rangle &=&\left\langle
E_{A0}^{+}E_{A0}^{-}E_{B1}^{+}E_{B1}^{-}\right\rangle =\left\langle
E_{A0}^{+}E_{A0}^{-}\right\rangle \left\langle
E_{B1}^{+}E_{B1}^{-}\right\rangle \\
&&+\left\langle E_{A0}^{+}E_{B1}^{+}\right\rangle \left\langle
E_{A0}^{-}E_{B1}^{-}\right\rangle +\left\langle
E_{A0}^{+}E_{B1}^{-}\right\rangle \left\langle
E_{A0}^{-}E_{B1}^{+}\right\rangle \\
&=&\left\langle I_{A0}\right\rangle \left\langle I_{B2}\right\rangle +\left|
\left\langle E_{A0}^{+}E_{B1}^{+}\right\rangle \right| ^{2},
\end{eqnarray*}
the former equality deriving from the property of the average of four
Gaussian random variables and the average $\left\langle
E_{A0}^{+}E_{B1}^{+}\right\rangle $ not contributing as may be realized. A
similar procedure may be used for $\left\langle I_{A2}I_{B0}\right\rangle $
whence eq.$\left( \ref{s11}\right) $ becomes 
\begin{equation}
R_{AB}=\left| \left\langle E_{A0}^{+}E_{B1}^{+}\right\rangle \right|
^{2}+\left| \left\langle E_{A1}^{+}E_{B0}^{+}\right\rangle \right|
^{2}+\left\langle I_{A1}I_{B1}\right\rangle .  \label{s14}
\end{equation}

This result differs from the quantum prediction got via the WW formalism, eq.%
$\left( \ref{e2}\right) ,$ asides from the irrelevant scale factor 2 (see
comment after eq.$\left( \ref{s5}\right) )$ due to the presence of the term $%
\left\langle I_{A1}I_{B1}\right\rangle .$ Now we argue that this term does
not contribute. In fact writing it in terms of fields we have 
\begin{equation}
\left\langle I_{A1}I_{B1}\right\rangle =\left\langle \left(
E_{A0}^{+}E_{A1}^{-}+E_{A1}^{+}E_{A0}^{-}\right) \left(
E_{B0}^{+}E_{B1}^{-}+E_{B1}^{+}E_{B0}^{-}\right) \right\rangle .  \label{s15}
\end{equation}
Performing the product there are 4 terms that we may calculate using again
the property of the product of four Gaussian variables. The former term
leads to a nul result, namely 
\begin{eqnarray*}
\left\langle E_{A0}^{+}E_{A1}^{-}E_{B0}^{+}E_{B1}^{-}\right\rangle
&=&\left\langle E_{A0}^{+}E_{A1}^{-}\rangle \langle
E_{B0}^{+}E_{B1}^{-}\right\rangle +\left\langle E_{A0}^{+}E_{B0}^{+}\rangle
\langle E_{A1}^{-}E_{B1}^{-}\right\rangle \\
+\left\langle E_{A0}^{+}E_{B1}^{-}\rangle \langle
E_{A1}^{-}E_{B0}^{+}\right\rangle &=&0,
\end{eqnarray*}
because all the averages are zero taking eqs.$\left( \ref{b3}\right) $ into
account. Similarly $\left\langle
E_{A1}^{+}E_{A0}^{-}E_{B1}^{+}E_{B0}^{-}\right\rangle =0.$ The two terms
remaining from eq.$\left( \ref{s15}\right) $ are complex conjugate of each
other so that we may write 
\begin{equation}
\left\langle I_{A1}I_{B1}\right\rangle =2Re\left\langle
E_{A0}^{+}E_{A1}^{-}E_{B1}^{+}E_{B0}^{-}\right\rangle .  \label{s16}
\end{equation}
If we evaluated this expectation using the Gaussian property as previously,
we would get 
\begin{eqnarray*}
\left\langle E_{A0}^{+}E_{A1}^{-}E_{B1}^{+}E_{B0}^{-}\right\rangle
&=&\left\langle E_{A0}^{+}E_{A1}^{-}\rangle \langle
E_{B1}^{+}E_{B0}^{-}\right\rangle +\left\langle E_{A0}^{+}E_{B1}^{+}\rangle
\langle E_{A1}^{-}E_{B0}^{-}\right\rangle \\
+\left\langle E_{A0}^{+}E_{B0}^{-}\rangle \langle
E_{A1}^{-}E_{B1}^{+}\right\rangle &=&\left\langle
E_{A0}^{+}E_{B1}^{+}\rangle \langle E_{A1}^{-}E_{B0}^{-}\right\rangle
+\left\langle E_{A0}^{+}E_{B0}^{-}\rangle \langle
E_{A1}^{-}E_{B1}^{+}\right\rangle ,
\end{eqnarray*}
the former term being nil. The latter two terms consist of a product of two
averages, or expectation values, each. Every average consists of a field
amplitude arriving at Alice times another amplitude arriving at Bob. In
these conditions we cannot ignore the space-time factors like $\exp \left[ i%
\mathbf{k\cdot r}-i\omega t\right] ,$ see eq.$\left( \ref{55}\right) ,$ that
would be different in the Alice and Bob beams, and uncorrelated. Therefore
it is plausible that the average over these phases should result in a null
value for the expectation and consequently the average eq.$\left( \ref{s16}%
\right) $ will be zero, so ending the proof that the term $\left\langle
I_{A1}I_{B1}\right\rangle $ of eq.$\left( \ref{s14}\right) $ does not
contribute to order $\left| D\right| ^{2}$. I point out that the condition
is quite different in the other terms of eq.$\left( \ref{s14}\right) $
because they involve absolute values, making the space-time phases
irrelevant in the average. Thus we get finally the prediction of our local
model for the coincidence detection rate 
\begin{equation}
R_{AB}=\left| \left\langle E_{A0}^{+}E_{B1}^{+}\right\rangle \right|
^{2}+\left| \left\langle E_{A1}^{+}E_{B0}^{+}\right\rangle \right| ^{2},
\label{s20}
\end{equation}
that agrees with the quantum prediction eq.$\left( \ref{e2}\right) ,$ except
for a scale factor 2 (see comment after eq.$\left( \ref{s5}\right) )$.
Indeed taking eqs.$\left( \ref{b3}\right) $ into account, we get the
coincidence detection rate 
\begin{equation}
R_{AB}=\frac{1}{2}\left| D\right| ^{2}\cos ^{2}(\theta -\phi ).  \label{s12}
\end{equation}

As a conclusion the results of our realistic model, eqs.$\left( \ref{s5}%
\right) $ and $\left( \ref{s12}\right) ,$ agree with the quantum predictions
eqs.$\left( \ref{3.3}\right) ,\left( \ref{3.4}\right) $ and $\left( \ref{3.7}%
\right) ,$ modulo an scaling parameter 1/2. Indeed the predictions of the
model, eqs.$\left( \ref{s5}\text{ }\right) $ and $\left( \ref{s12}\right) ,$
have the form of eq.$\left( \ref{pred}\right) $ and therefore violate a Bell
inequality.

It is interesting for the picture of the phenomenon, to be discussed in the
next subsection, to write eq.$\left( \ref{s20}\right) $ in terms of
intensities, rather than amplitudes. This may be achieved taking eqs.$\left( 
\ref{s11}\right) $ and $\left( \ref{s13}\right) $ into account and
neglecting the term $\left\langle I_{A1}I_{B1}\right\rangle $ which is of
order $\left| D\right| ^{4}$. Thus we get 
\begin{equation}
R_{AB}=\left\langle (I_{A0}-\left\langle I_{A0}\right\rangle
)I_{B2}\right\rangle +\left\langle (I_{B0}-\left\langle I_{B0}\right\rangle
)I_{A2}\right\rangle  \label{s19}
\end{equation}

\subsection{A physical picture of the experiments}

Our model, resting upon the WW formalism of quantum optics, provides a
picture quite different from the one suggested by the HS formalism in terms
of photons. We do not assume that the photocount probability within a time
window factorizes so that there is a \textit{small} probability, of order $%
\left| D\right| ^{2},$ that the pumping laser produces in the crystal an
``entangled photon pair'' within a given time window, and then there is a
detection probability of order\textit{\ unity} conditional to the photon
pair production. (The latter probability is identified with the detection
efficiency). Furthermore the concept of photon does not appear at all, but
there are \textit{continuous fluctuating fields including a real ZPF}
arriving at the detectors, that are activated when the fluctuations are big
enough.

It is interesting to study more closely the ``quantum correlation''
qualified as strange from a classical point of view because it is a
consequence of the phenomenon of entanglement. The origin is the correlation
between the signal $I_{B1}$ produced by the action of the laser on the
crystal and the part $I_{A0}$ of the ZPF that entered in the crystal. This
correlation is essential for the large value of the coincidence rate eq.$%
\left( \ref{s12}\right) .$ In fact the detection probability by Bob, that we
might expect to be simply $\left\langle I_{B1}\right\rangle ,$ is enhanced
by the correlation deriving from the fact that \textit{the same normal mode
appears in both radiation fields, }$E_{A0}^{+}$\textit{\ and }$E_{B1}^{+}$
see eq.$\left( \ref{b3}\right) $. And similarly for $E_{A1}^{+}$ and $%
E_{B0}^{+}.$ More specifically with reference to eq.$\left( \ref{s19}\right) 
$ I point out that the ensemble average of $I_{A0}-\left\langle
I_{A0}\right\rangle $ being zero, \textit{only the fluctuations} are
involved in the enhancement of detection probability by Bob. Indeed in a
fluctuation such that $I_{A0}-\left\langle I_{A0}\right\rangle >0$ $(<0)$ eq.%
$\left( \ref{b3}\right) $ shows that also $I_{B1}>0$ $(<0)$ whence the
average of their product is always positive, that is $\left\langle
I_{B1}\left( I_{A0}-\left\langle I_{A0}\right\rangle \right) \right\rangle
>0.$ And similarly for the fluctuations of (Bob) term $I_{B0}-\left\langle
I_{B0}\right\rangle $ that enhance the detection probability of Alice due to 
$I_{A1}$. This leads to an interesting interpretation of entanglement: 
\textit{it is a correlation between }\textbf{fluctuations}\textit{\
involving the vacuum fields}.

\subsection{Conclusions}

I conclude that the hypothesis that the quantum vacuum fields are real
allows a concept of locality weaker than the fulfillement of the Bell
inequalities. Indeed in our model the signal fields (accompanied by vacuum
fields) travel causally from the source (that is the laser pumping beam and
the nonlinear crystal followed by a beam-splitter and other devices) to the
detectors. Thus I claim that the model of this paper is local. On the other
hand the results for the single and coincidence detection probabilities
within a time window, eqs.$\left( \ref{s5}\right) $ and $\left( \ref{s12}%
\right) ,$ violate a Clauser-Horne (Bell) inequality eq.$\left( \ref{CH}%
\right) $. Therefore it is possible to explain the violation of Bell
inequalities in entangled photon experiments without any conflict with
relativistic causality.

\end{document}